\documentclass[prb,aps,amssymb,showpacs,twocolumn]{revtex4}

\usepackage{amsmath}
\usepackage{amssymb}
\usepackage{amsthm}
\usepackage{amsfonts}

\usepackage{enumerate}
\usepackage{latexsym}
\usepackage[dvips]{graphicx}

\newcommand{\beq}{\begin{equation}}
\newcommand{\eneq}{\end{equation}}
\newcommand{\bed}{\begin{displaymath}}
\newcommand{\ened}{\end{displaymath}}
\input{epsf}

\def\simge{\mathrel{\rlap{\raise 0.511ex \hbox{$>$}}{\lower 0.511ex
   \hbox{$\sim$}}}}
\def\simle{\mathrel{\rlap{\raise 0.511ex \hbox{$<$}}{\lower 0.511ex
   \hbox{$\sim$}}}}

\begin{document}

\tolerance 10000


\title{Spin Density Wave and D-Wave Superconducting Order Parameter ``Coexistence''}

\author { Zaira Nazario$^\dagger$ and 
David I. Santiago$^{\dagger, \star}$ }

\affiliation{$\dagger$ Department of Physics, Stanford University,
             Stanford, California 94305 \\ $\star$ Gravity Probe B
             Relativity Mission, Stanford, California 94305}
\begin{abstract}

\begin{center}

\parbox{14cm}{ We study the properties of a spin-density-wave
antiferromagnetic mean-field ground state with d-wave superconducting
(DSC) correlations. This ground state always gains energy by Cooper
pairing. It would fail to superconduct at half-filling due to the
antiferromagnetic gap although its particle-like excitations would be
Bogolyubov-BCS quasiparticles consisting of coherent mixtures of
electrons and holes. More interesting and relevant to the
superconducting cuprates is the case when antiferromagnetic order is
turned on weakly on top of the superconductivity. This would
correspond to the onset of antiferromagnetism at a critical doping. In
such a case a small gap proportional to the weak antiferromagnetic gap
opens up for nodal quasiparticles, and the quasiparticle peak would be
discernible. We evaluate numerically the absorption by nodal
quasiparticles and the local density of states for several ground
states with antiferromagnetic and d-wave superconducting
correlations. }

\end{center}
\end{abstract}


\maketitle

\section{Introduction}

Ever since the discovery of high temperature
superconductivity\cite{htcexp} it was proposed that the
superconducting correlations might already exists in the
antiferromagnetic Mott insulator\cite{rvb}. The origin of the
superconducting correlations was ascribed to the large Coulombic
interactions in the undoped materials. The only other large energy
scale in the materials is phononic\cite{zx1}.

While the microscopic origin of superconductivity remains a matter of
debate\cite{rvb2, goss1, goss2, zx2}, there is growing experimental
evidence that the quasiparticles are Bogolyubov-BCS
quasiparticles. Bending back of photoemission bands\cite{campu1}
fits quantitatively the BCS-Bogolyubov model\cite{bcs, bogo1}. Scanning
tunneling microscopy finds coherent quasiparticles that disperse as a
coherent mixture of particles and holes\cite{seamus,seamus2}. The particle and
hole amplitudes in these experiments and in inverse photoemission
experiments\cite{seamus2, ding} fit accurately to the theoretical
Bogolyubov-BCS values calculated from the dispersion and gap measured
in the normal and superconducting materials respectively.

\begin{figure}
  \begin{center}
    \begin{tabular}{cc}
      \resizebox{8cm}{!}{\rotatebox{270}{\includegraphics{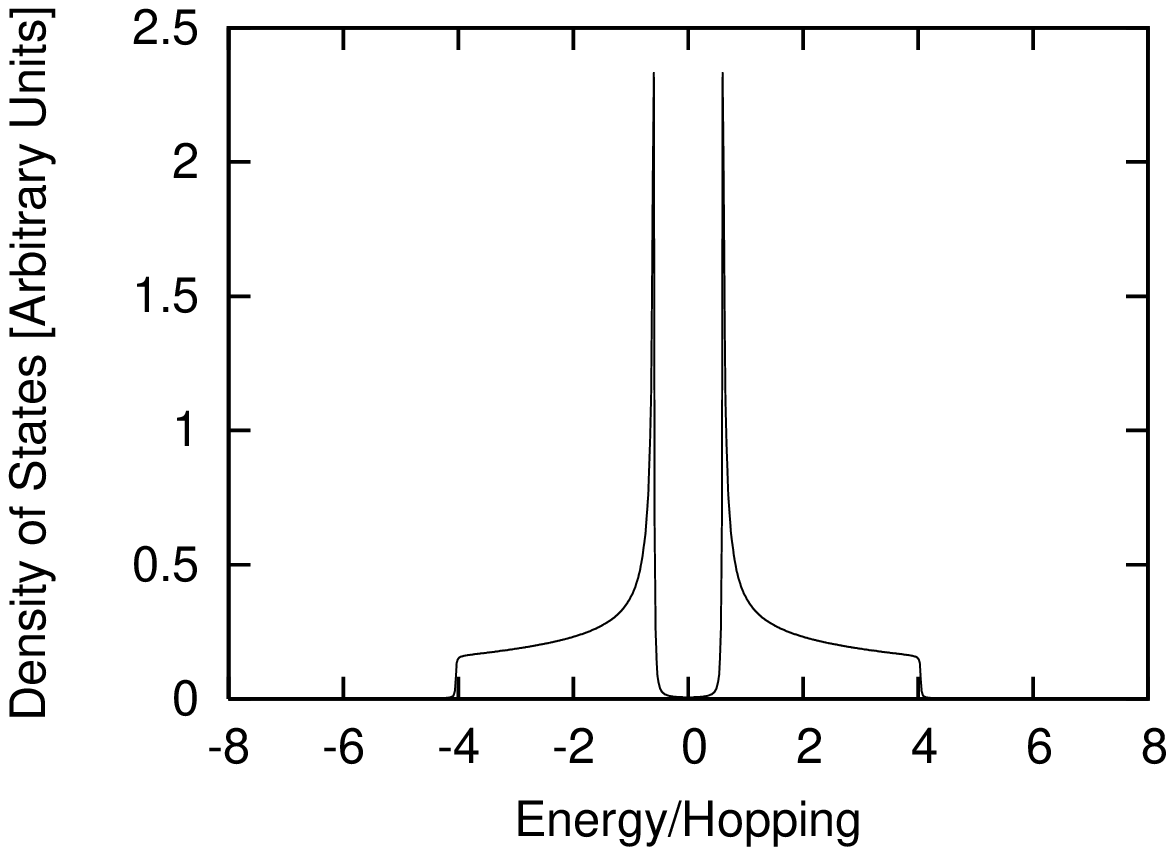}}} \\
      \resizebox{8cm}{!}{\rotatebox{270}{\includegraphics{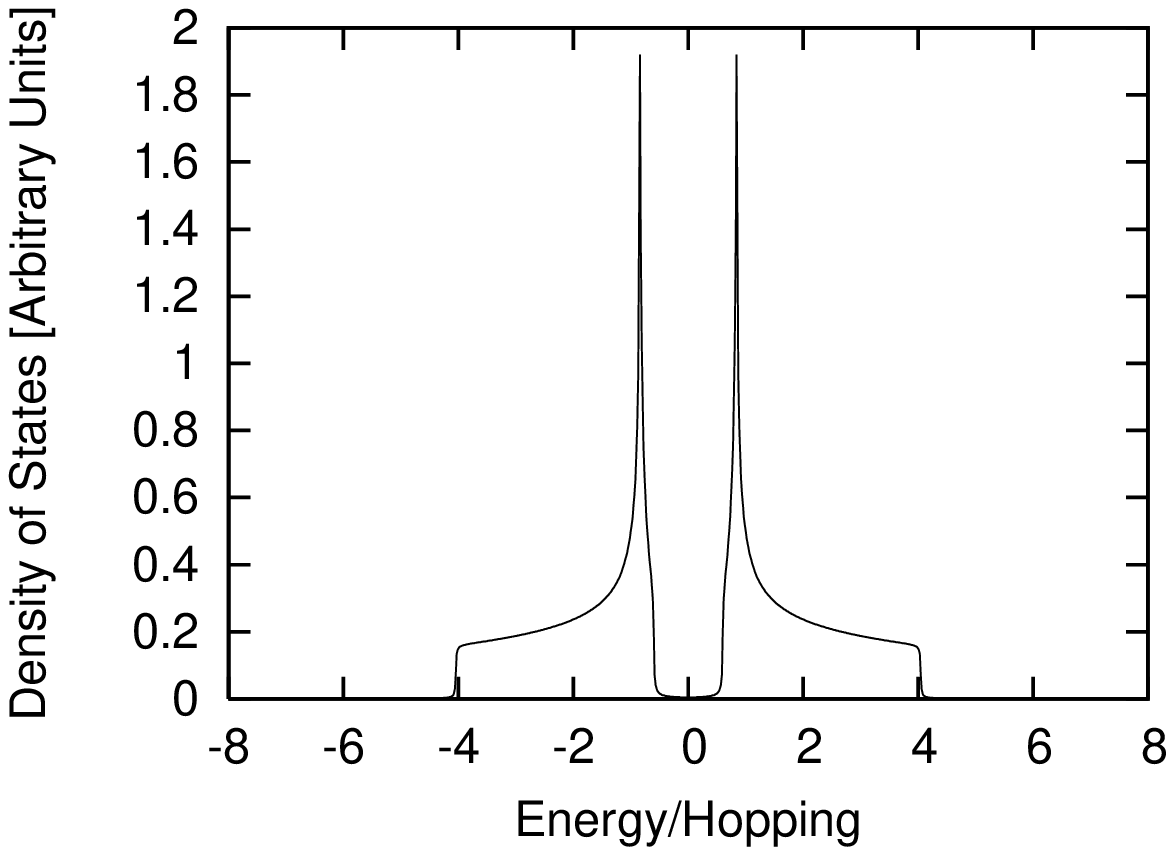}}}
    \end{tabular}
    \caption{Spectral Function for the SDW ground state
    without (1a) and with (1b) superconducting correlations built
    in.}
  \end{center}
\end{figure}

Regardless of whether the origin of superconducting correlations is
exotic Coulombic physics or some more conventional mechanism, it is
clear that the cuprates are BCS paired superconductors. This does not
mean that the Coulomb interactions do not matter. Rather, the
interesting and contradictory physics for underdoped materials is the
result of Coulomb degradation of the superfluid density\cite{rvb,
rvb2, goss1, goss2} and order parameter competition between
superconductivity and correlated electron ground
states\cite{ordercomp, ddw}. The degradation of the superfluid density
leads to suppressed $T_c$ due to a phase instability of the
superconducting order parameter\cite{uemura, ong, rvb, kiv,
vort}. There are several Coulomb stabilized competing ground states
such as orbital antiferromagnetism\cite{ddw}, stripe or charge density
wave ground states\cite{kiv2} and perhaps electronic liquid crystal
phases\cite{kiv3}. Regardless of which of these competing ground
states are realized, there is strong experimental evidence for
incommensurate electronic ordering\cite{neutron}, either static or
incipient. The evidence seems more consistent with charge-density-wave
or stripe order.

In the present work we will study the physics of an antiferromagnet
with a strong d-wave Cooper pairing interaction. We do not speculate as
to the origin of this superconducting interaction except to point out
that in such a model it competes with the Coulombic antiferromagnetic
physics. Both the superconductor and the antiferromagnet are studied
in the mean field approximation. While one can doubt the validity of
such an approximation at a phase transition point, it will be
qualitatively correct within the ordered phases. 

Cooper pairing leading to a BCS ground state is an instability of a
Fermi liquid ground state. In this study we apply the BCS
approximation to a spin density wave (SDW) insulating ground state as
it exists in the cuprates at half filling. The resulting ground state 
{\it has Cooper pairing} yet it {\it fails to superconduct}
due to the SDW insulating gap. Next we will review some well known
facts in order to understand how a state with Cooper pairs does not
superconduct. Before doing so we emphasize that this only happens as a
consequence of having a {\it completely filled} insulating band.

\begin{figure}
  \resizebox{7cm}{!}{\includegraphics{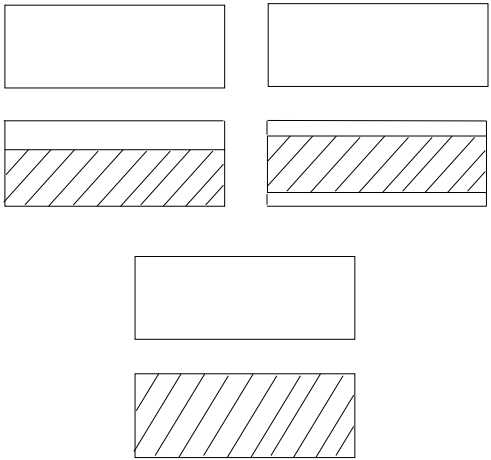}} 
  \caption{In the upper left side we draw two bands separated by a
  gap, with the lower band partially filled appropriate to a metal or
  superconductor. In the upper right side we draw the situation
  encountered for the electron fluid under the action of an electric
  field. The lower sketch illustrates the situation appropriate to an
  insulator, where the lower band is completely filled, making
  conduction impossible regardless of Cooper correlations.}
\end{figure}

When an electric field is applied to a metal, it conducts
dissipatively. The way this happens is that the center of mass of the
Fermi sea gets displaced upward in the unfilled metallic
band\cite{mott-jones}. Ohmic dissipation occurs because newly filled
electronic states at the top of the Fermi sea get scattered into newly
empty electronic states at the bottom of the Fermi sea due to the
lack of rigidity of the Fermi liquid ground state (see figure
2). When there are Cooper correlations, the electron liquid gets
displaced upward in the band too, but as long as the displacement in
energy within the band is less than the superconducting gap, Cooper
pair correlations make the electron liquid rigid, thus preventing
scattering and dissipation. For the case of an SDW ground state at half
filling with Cooper pairing correlations there is no superconductivity
as the electron fluid {\it cannot} move upward in the band for the
band is full and {\it there are no electronic states} to be filled
unless one excites across the insulating gap and into the conduction
band (see figure 2).

That the SDW insulating ground state with d-wave pairing interactions
has Cooper pairing in the ground state can be seen from figure 1. In
figure 1a we plot the spectral function for the SDW ground state with
no superconducting correlations. In figure 1b we plot the spectral
function for the SDW ground state with superconducting
correlations. In the ground state with both superconductivity and
antiferromagnetism, the separation between the coherence peaks is
bigger as it gets contributions from both the SDW and superconducting
gap. A prediction of this model is that the quasiparticles will be
coherent with an electron and a hole component in agreement with the
BCS-Bogolyubov model.

The SDW ground state with d-wave Cooper pairing (SDW-DSC) will become
superconducting when doped. At the mean-field level, without worrying
about self consistency, the chemical potential will jump to the
appropriate band and there will be a low superfluid density
superconductor. Whether this physics is correct for the cuprates is
controversial. There is experimental evidence for the chemical
potential staying pinned at midgap due to spectral redistribution of
states toward midgap states\cite{optics}. There is also experimental
evidence for chemical potential shifts in the cuprates, in the same
way as in regular semiconductor materials\cite{kastner}. Independently
of whether the SDW-DSC ground state has chemical potential shifts or
not, the physics of an insulator with Cooper pairing correlations is
interesting. For our study we have the cuprates in mind. For these
materials, some phenomenology of this form seems to apply\cite{rvb},
but it would be interesting if this physics were to be realized in nature
irrespective of the cuprate problem.

In the present work we will flip the problem around. We will start
with a d-wave superconductor (DSC) and begin turning on SDW
antiferromagnetic order on top of the superconductivity. In this
limit, the complications mentioned in the previous paragraph are nonexistent. 
A slow turning on of SDW order on top of the
superconductivity will show up as a shift of the antinodal gap and a
gapping of the nodal quasiparticles. The latter should be a signal
much easier to pick out than the gap shift. The gapping of the nodal
quasiparticles is not a unique prediction of antiferromagnetic
ordering on top of the superconductivity, as such a gapping can be
produced by disorder. On the other hand, the coherence of the gapped
``nodal'' quasiparticles would be nonexistent for a disordered gap and
is thus a unique signature of antiferromagnetic ordering developing on
top of the superconductivity. Therefore, if a quasiparticle peak is
discernible, and the broadening is less than the disorder-induced
broadening ($\simge \hbar^2/2m\Delta x^2 \simeq 350$ meV for $\Delta
x \sim 1$ nm, appropriate to the cuprates), then the gap is a long
range ordered gap and not a disordered gap. Another unique signature
of an SDW gap is that the gap will open exactly at the doping where
the antiferromagnetism starts.

\begin{widetext}
\begin{figure}
  \begin{tabular}{ccc}
      \resizebox{7.8cm}{!}{\rotatebox{270}{\includegraphics{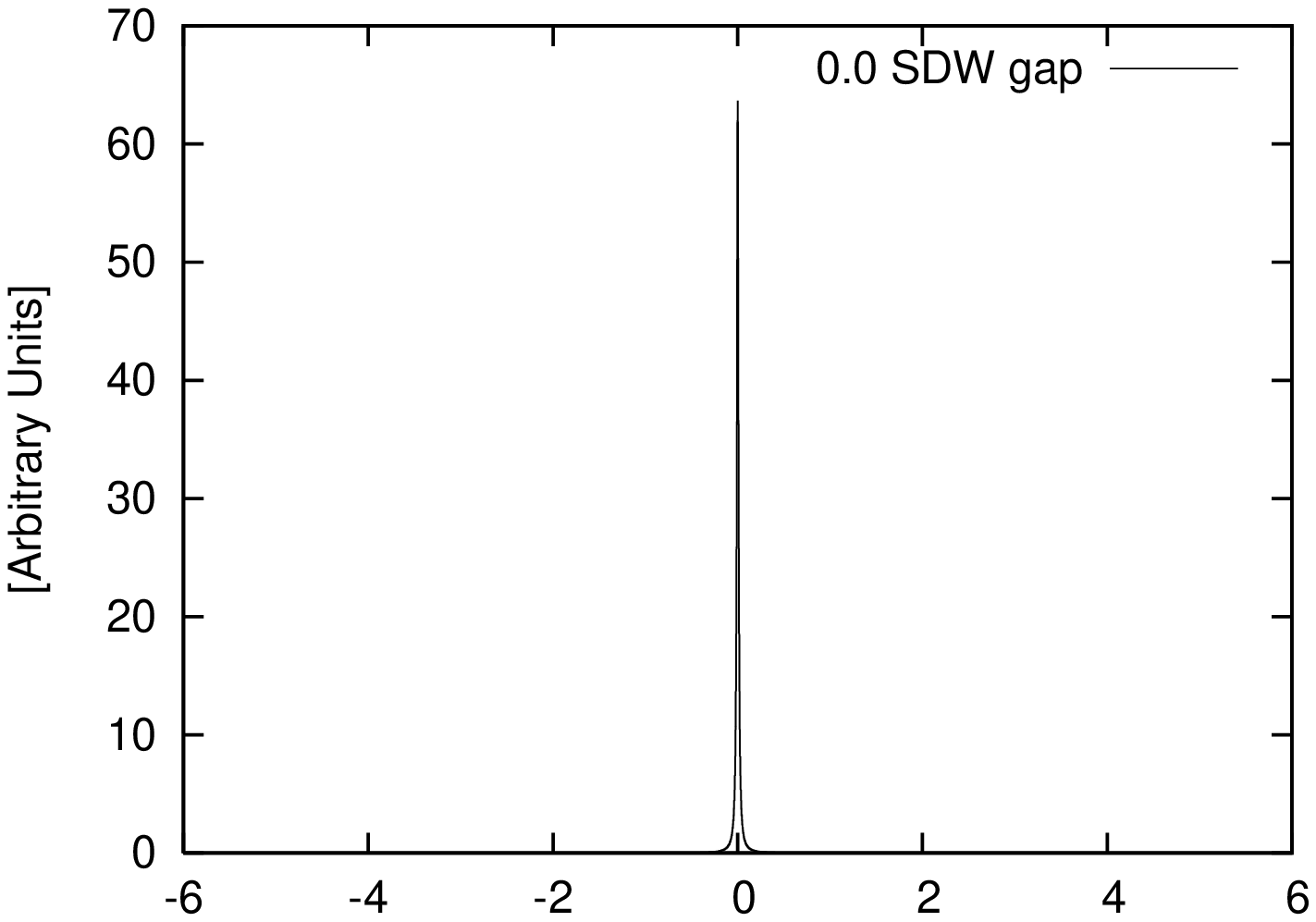}}} &
      \resizebox{7.8cm}{!}{\rotatebox{270}{\includegraphics{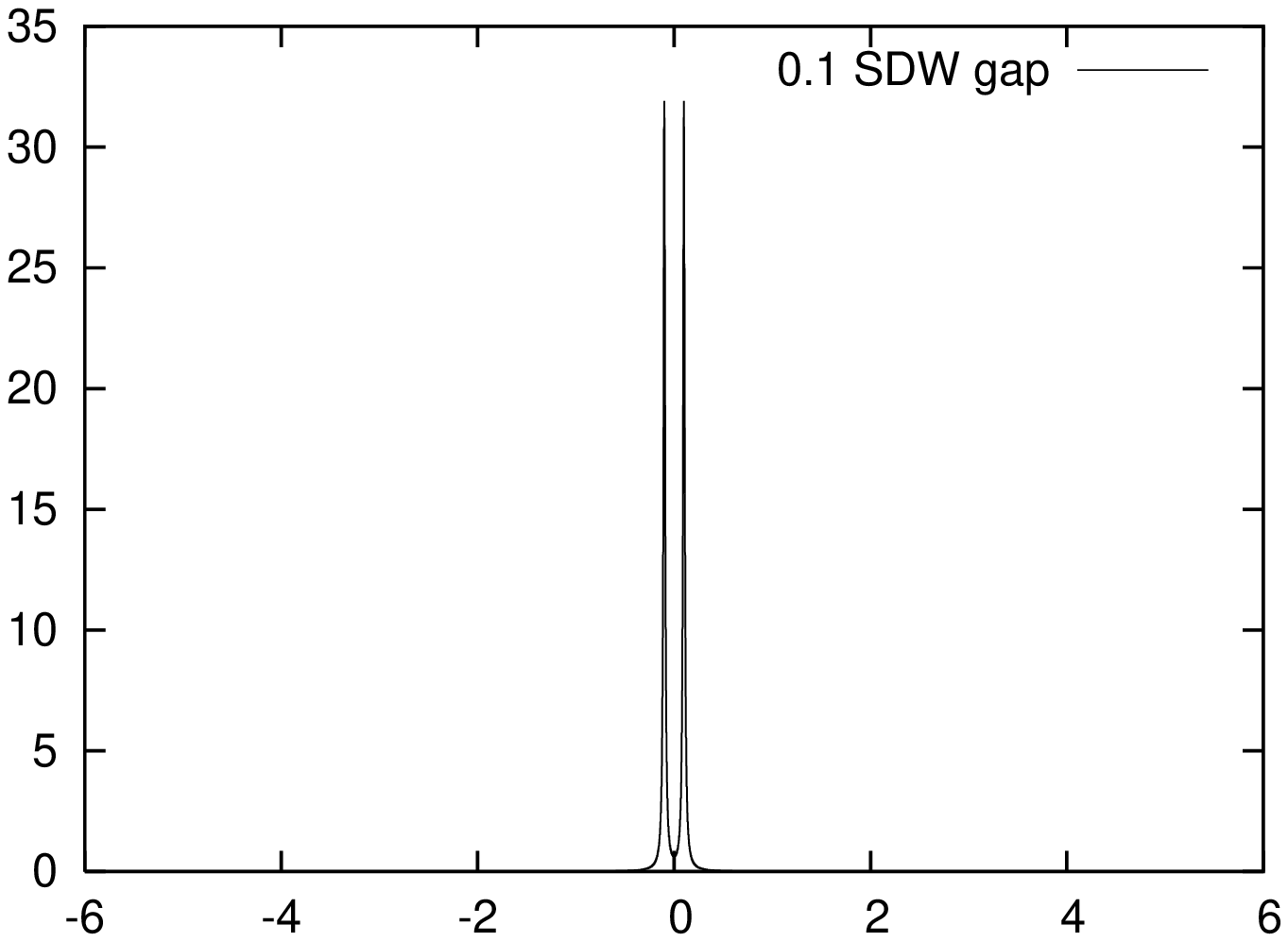}}}\\
      \resizebox{7.8cm}{!}{\rotatebox{270}{\includegraphics{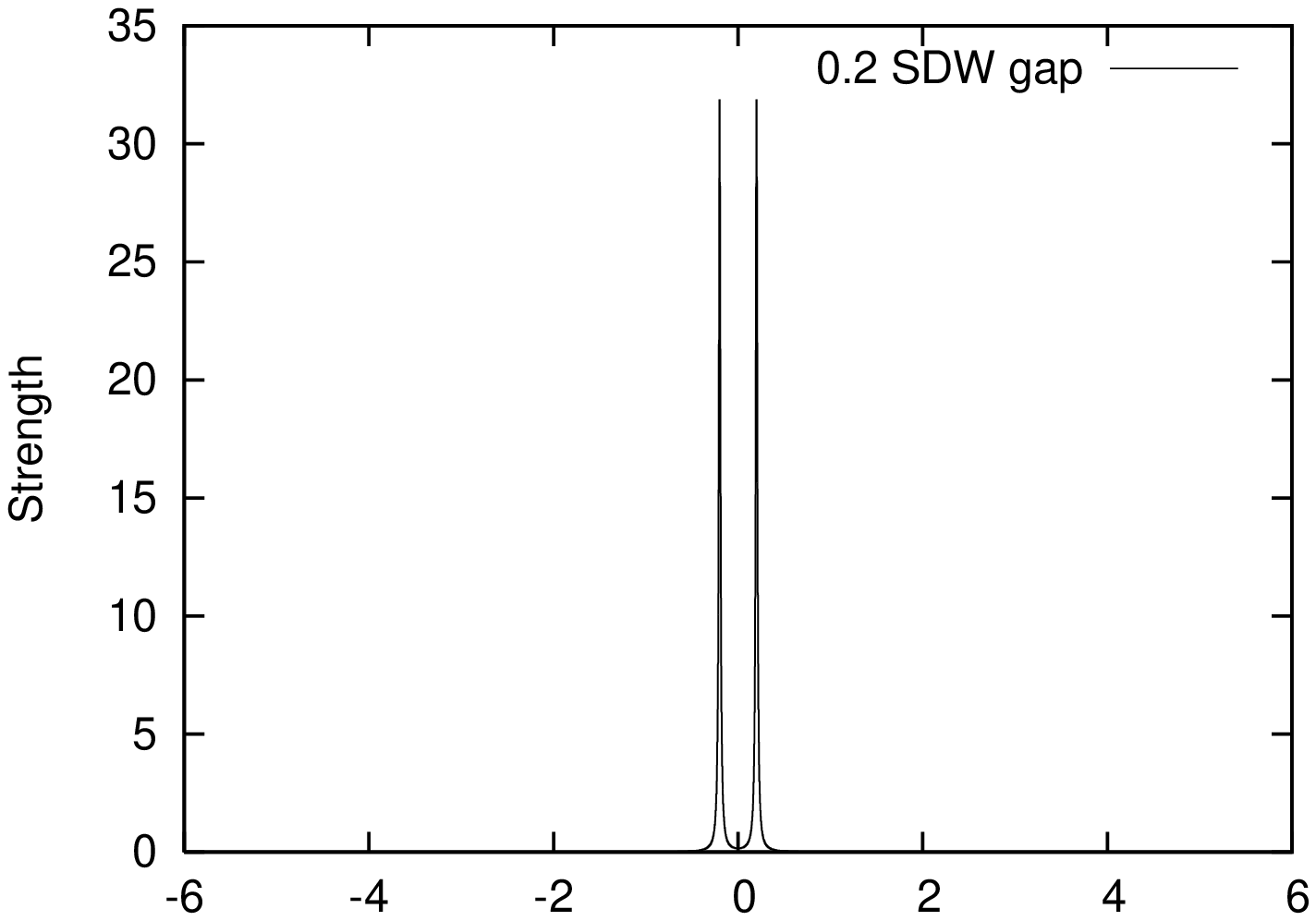}}}&
      \resizebox{7.8cm}{!}{\rotatebox{270}{\includegraphics{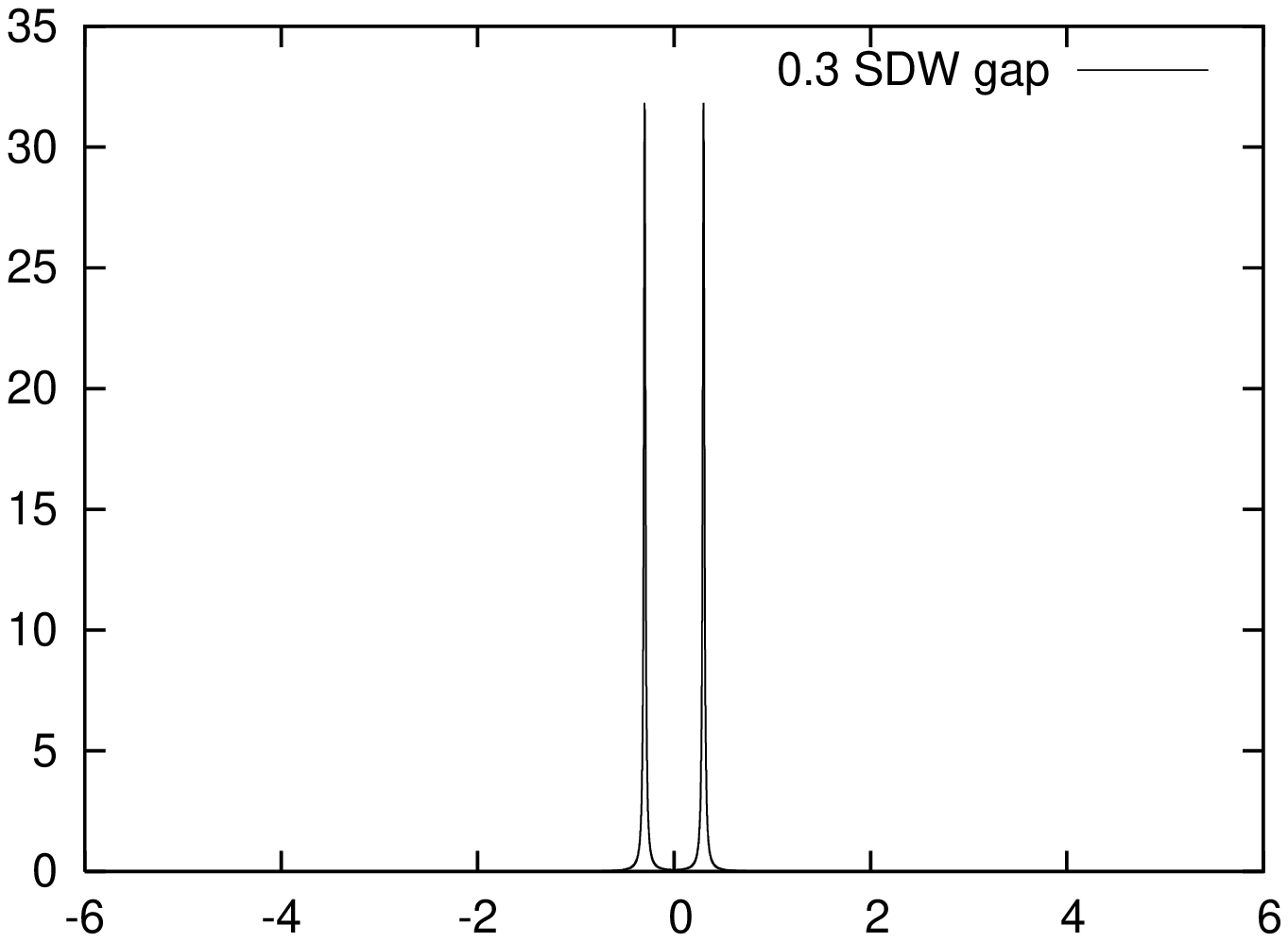}}}\\
      \resizebox{7.8cm}{!}{\rotatebox{270}{\includegraphics{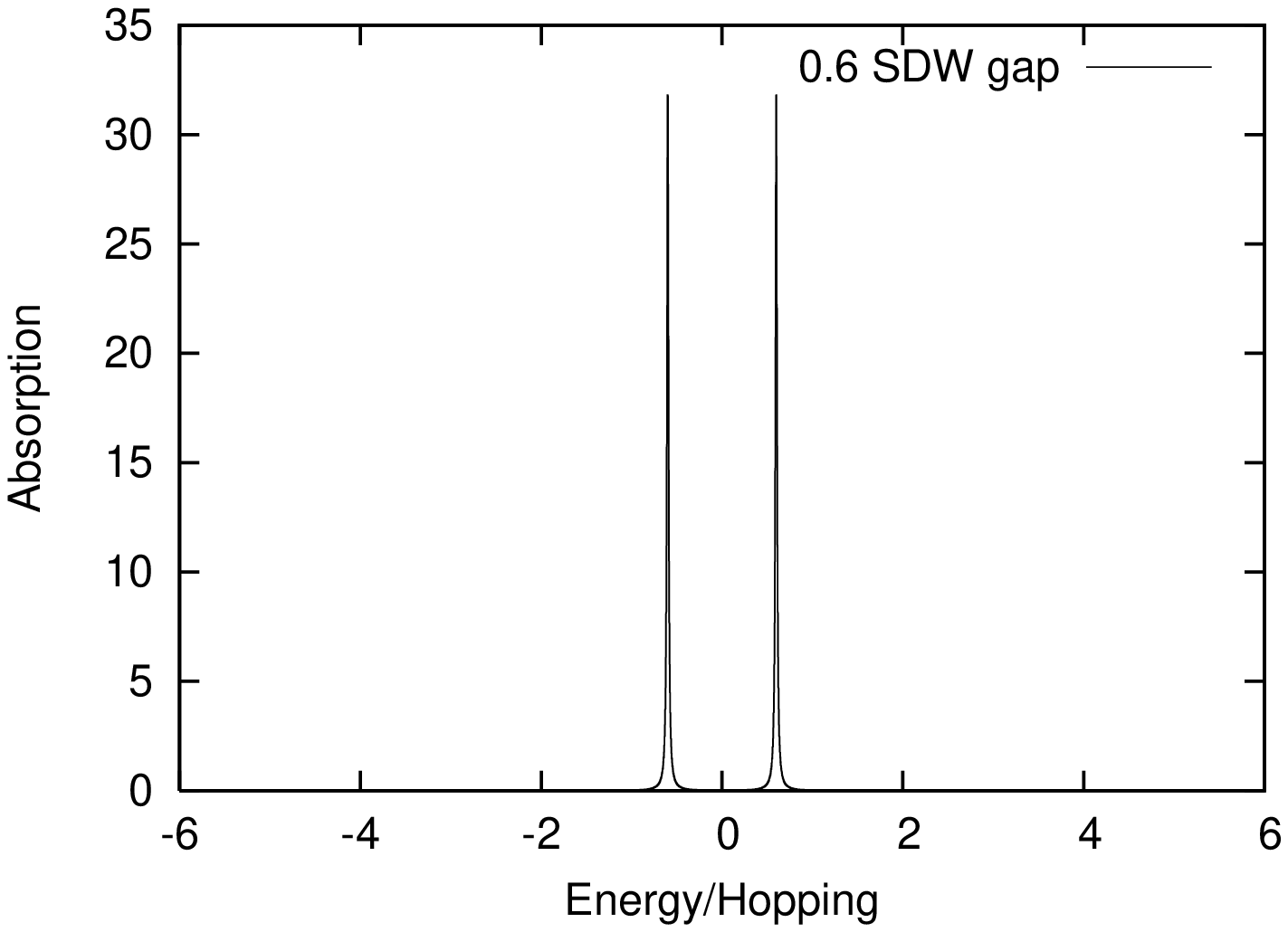}}}&
      \resizebox{7.8cm}{!}{\rotatebox{270}{\includegraphics{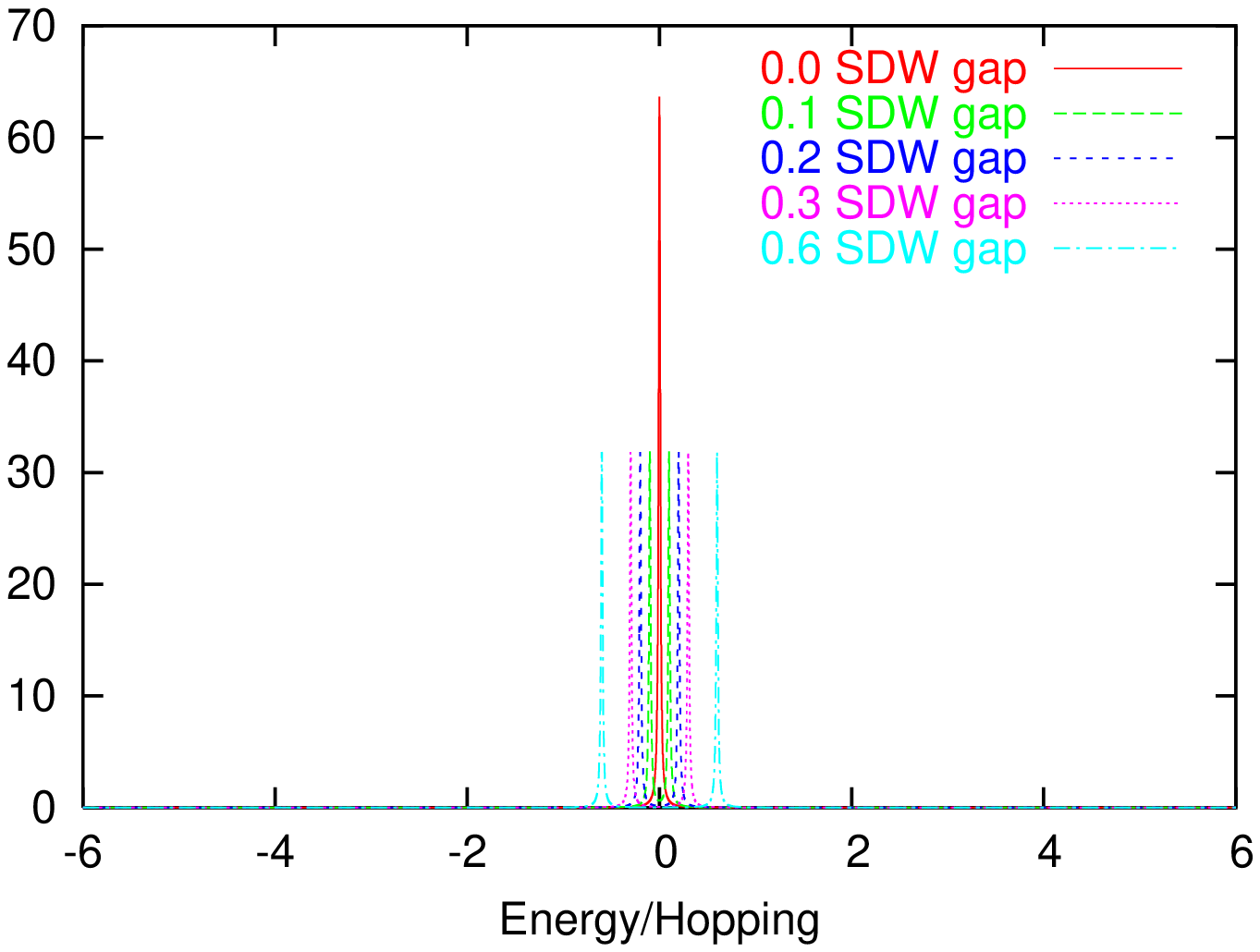}}}
  \end{tabular}
  \caption{Gapping of the nodal quasiparticles pole as the SDW order
  develops.}
\end{figure}
\end{widetext}

There are experimental suggestions of antiferromagnetism competing
with superconductivity in the deep underdoped regime in the
cuprates. For example, measurements show the nodal quasiparticle peaks
surviving right up to the doping where antiferromagnetism starts. The
spectral weight of such peaks diminishes with decreasing doping,
consistent with spectral weight being robbed from the superconducting
long range order by a competing long range order such as
antiferromagnetism\cite{zx4}. If one looks in the
antiferromagnetically ordered dopings, there are experimental
suggestions of a competing order parameter that conducts
efficiently. Most strikingly, there are measurements of metallic
conduction even below the Neel ordering temperature\cite{ando}.

The gapping of the nodal quasiparticles pole as the SDW order develops
on top of the superconductivity is shown in figure 3 for different
values of the SDW gap. The reason we only have a quasiparticle sharp
pole is that we have not modeled the realistic electronic self
energies relevant to the cuprates as they are irrelevant to the 
point 

\begin{widetext}
\begin{figure}
  \begin{center}
    \begin{tabular}{ccc}
      \resizebox{7.8cm}{!}{\rotatebox{270}{\includegraphics{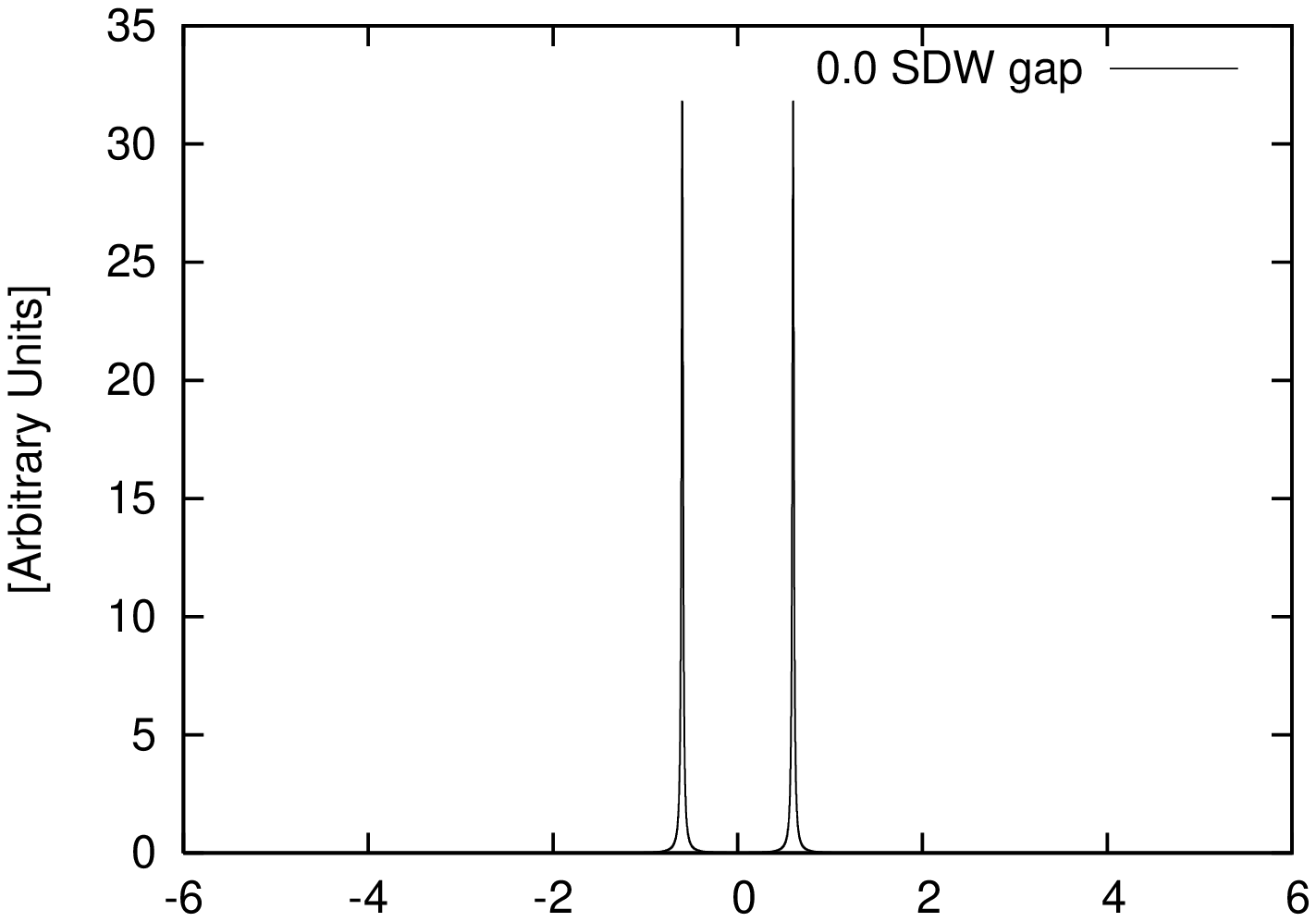}}} &
      \resizebox{7.8cm}{!}{\rotatebox{270}{\includegraphics{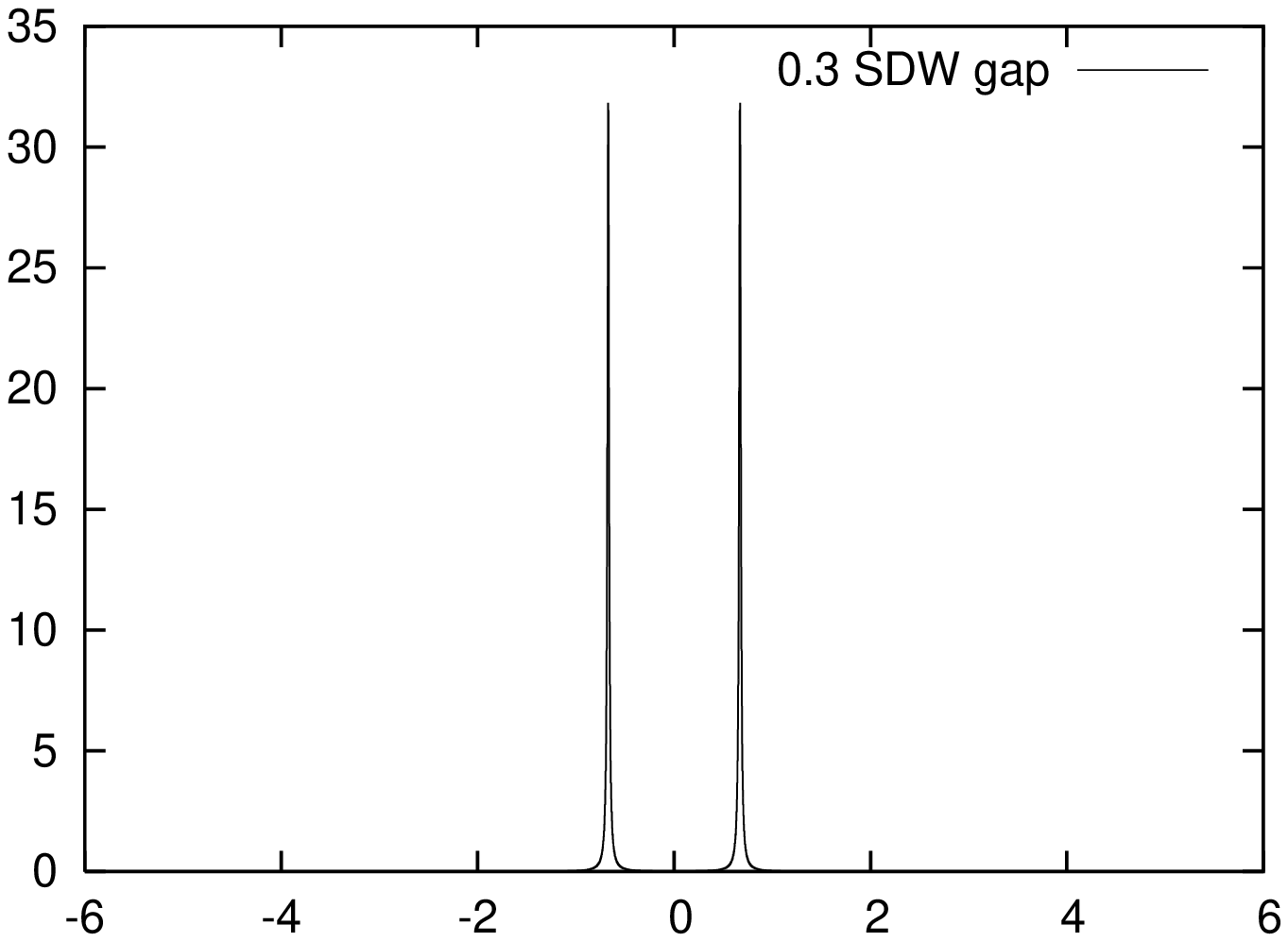}}}\\
      \resizebox{7.8cm}{!}{\rotatebox{270}{\includegraphics{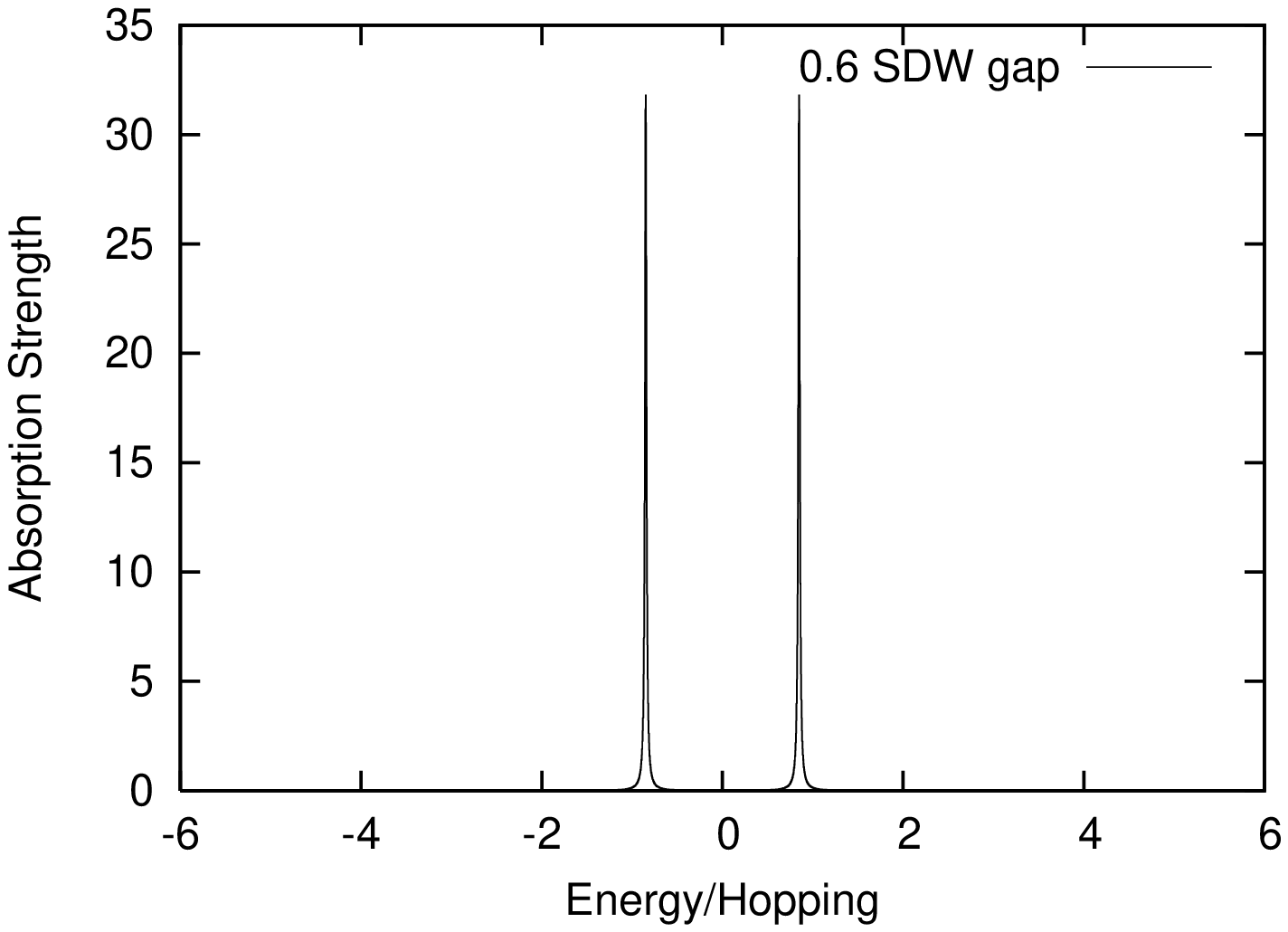}}}&
      \resizebox{7.8cm}{!}{\rotatebox{270}{\includegraphics{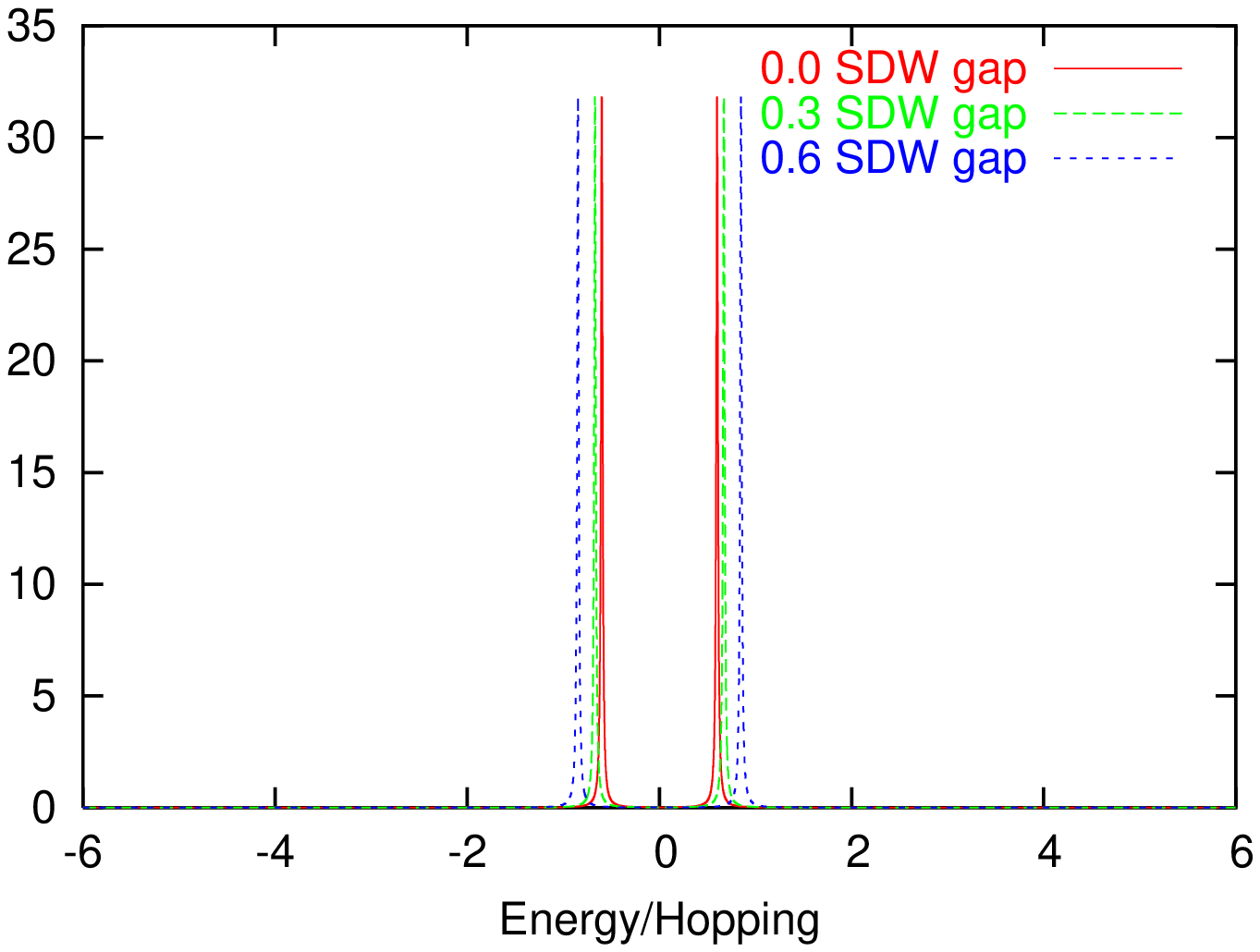}}}\\
    \end{tabular}
    \caption{Shift of the antinodal gap as the SDW order develops.}
  \end{center}
\end{figure}
\end{widetext}

\noindent  of principle we are making. Their only effect will be to 
broaden the quasiparticle peaks and add an incoherent background with the
phenomenological features. In figure 4 we plot the shift of the
antinodal gap as the SDW gap turns on. In figure 5 we plot the
spectral density of states in a d-wave superconductor as the SDW gap
is turned on. The superconductor with no SDW gap does not have a true
gap because of its d-wave symmetry. This is seen in the familiar
V-shaped collapse at zero energy. As the SDW gap is turned on, we see
the V-shape flatten and expand as a signature of the opening of the
antiferromagnetic gap.

\section{Hubbard Model with D-Wave Attractive Interactions}

For the cuprate problem, the two large effects are the
antiferromagnetic, or Coulombic, physics and the strong
superconductivity. Hence we will start from a phenomenological
Hamiltonian which is a Hubbard model with a d-wave electronic
interaction. This interaction will give rise to d-wave
superconductivity when we make the mean-field BCS approximation. The
Hamiltonian is

\begin{align} 
\mathcal{H} &= \nonumber \sum_{\vec{k},\sigma} (\epsilon_{\vec{k}} - \mu) \;
c_{\vec{k},\sigma}^\dagger c_{\vec{k},\sigma} + \frac{U}{N}
\sum_{\vec{k}_1,\vec{k}_2,\vec{q}} c_{\vec{k}_1,\uparrow}^\dagger
c_{\vec{k}_1 +\vec{q},\uparrow} c_{\vec{k}_2+\vec{q},
\downarrow}^\dagger c_{\vec{k}_2,\downarrow} \\
&+ \sum_{\vec{k}_1,\vec{k}_2} V(\vec{k}_1,\vec{k}_2) \;
c_{\vec{k}_1,\uparrow}^\dagger c_{-\vec{k}_1,\downarrow}^\dagger
c_{-\vec{k}_2,\downarrow} c_{\vec{k}_2,\uparrow}
\end{align}

\noindent where $c_{\vec{k},\sigma}^\dagger, c_{\vec{k},\sigma}$ are
the electronic creation and destruction operators with momentum
$\vec{k}$ and spin $\sigma$, $\epsilon_{\vec{k}}$ is the kinetic
energy, $\mu$ the chemical potential, and $U$ is the Hubbard
repulsion. We are working in a spatial lattice with $N$ sites. The
last term is an electronic interaction chosen in the reduced BCS
form\cite{bcs}, which will be used to stabilize superconductivity. In
order to have d-wave superconductivity we choose
$V(\vec{k}_1,\vec{k}_2) = V_0 \; (cos{k_{1x}} - cos{k_{1y}}) \;
(cos{k_{2x}} - cos{k_{2y}})$. This phenomenological Hamiltonian can
have a mean-field SDW ground state and a mean-field DSC ground
state. It can be used to study the turning on of DSC correlations on
top of an SDW ground state, or the turning on of SDW order on top of
the superconductivity. 

\begin{widetext}
\begin{figure}
  \begin{center}
    \begin{tabular}{ccc}
      \resizebox{7.8cm}{!}{\rotatebox{270}{\includegraphics{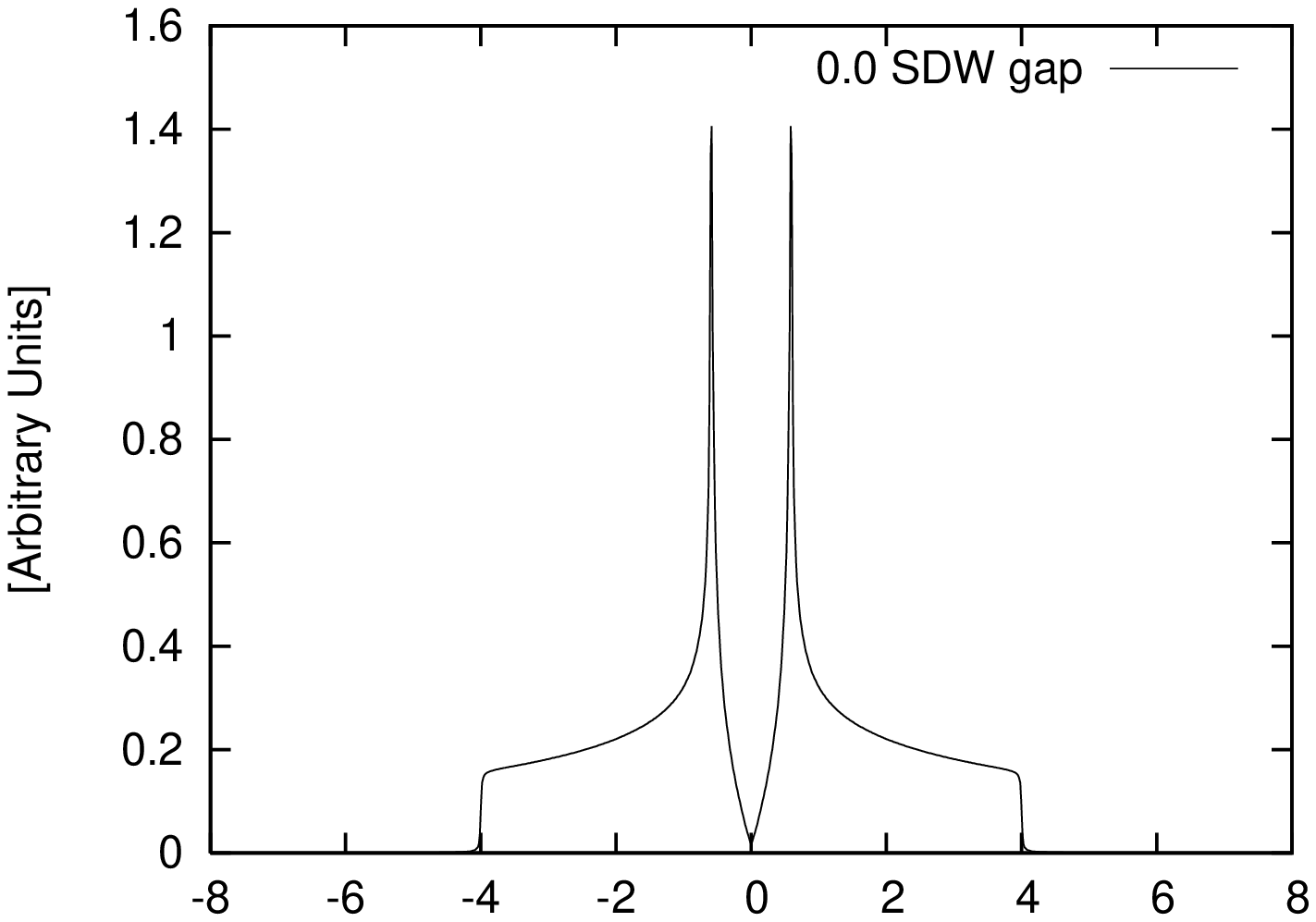}}} &
      \resizebox{7.8cm}{!}{\rotatebox{270}{\includegraphics{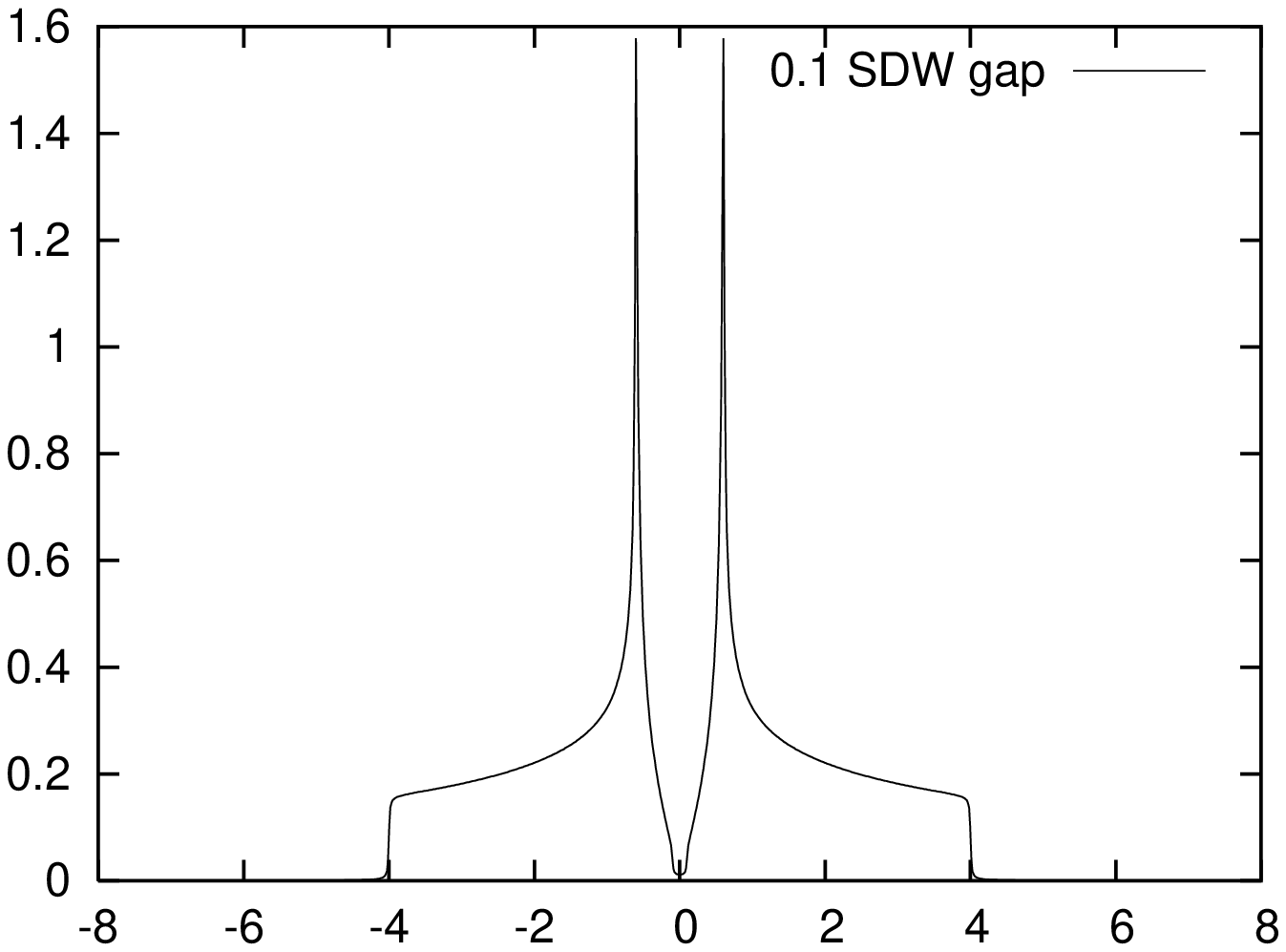}}}\\
      \resizebox{7.8cm}{!}{\rotatebox{270}{\includegraphics{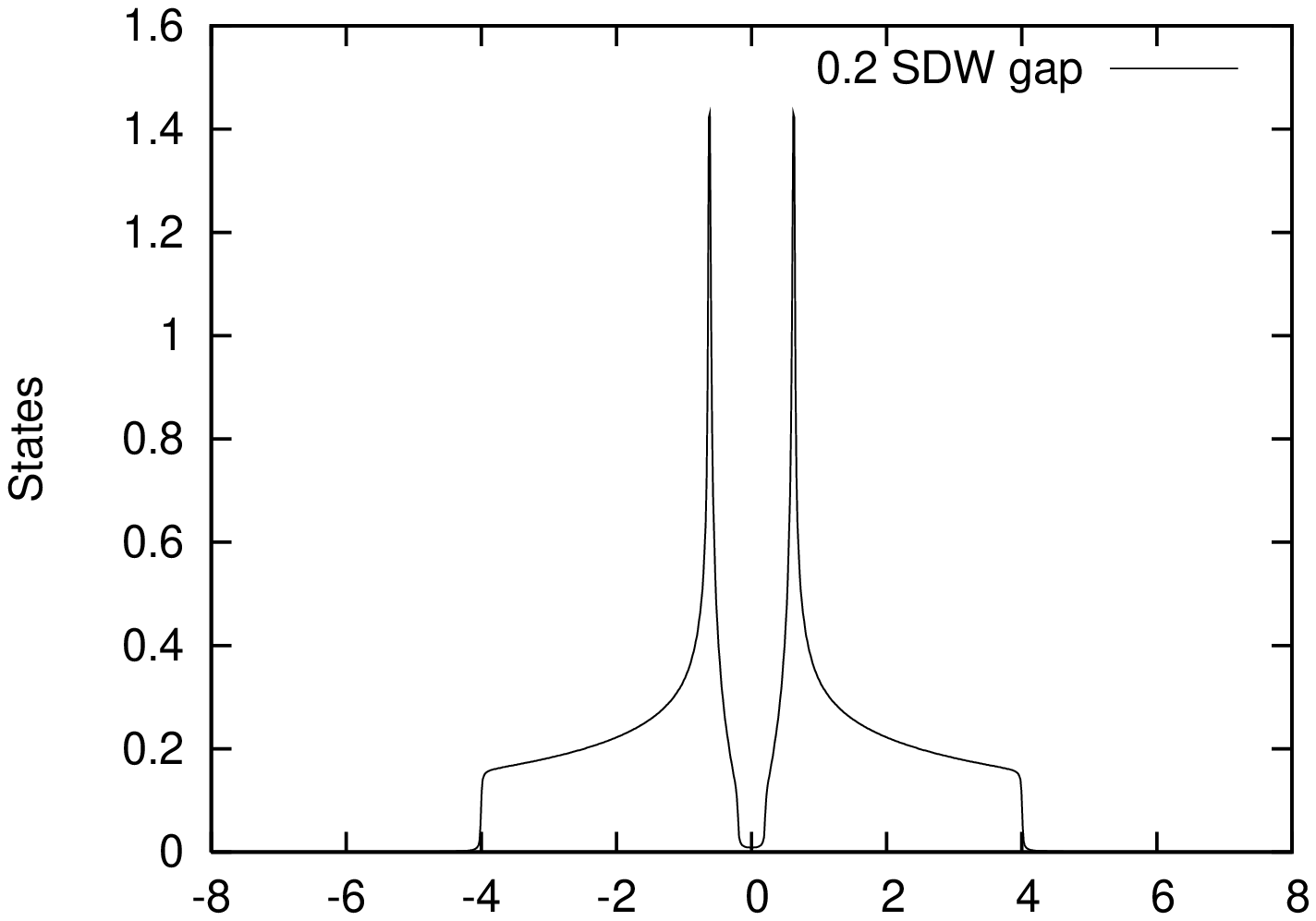}}}&
      \resizebox{7.8cm}{!}{\rotatebox{270}{\includegraphics{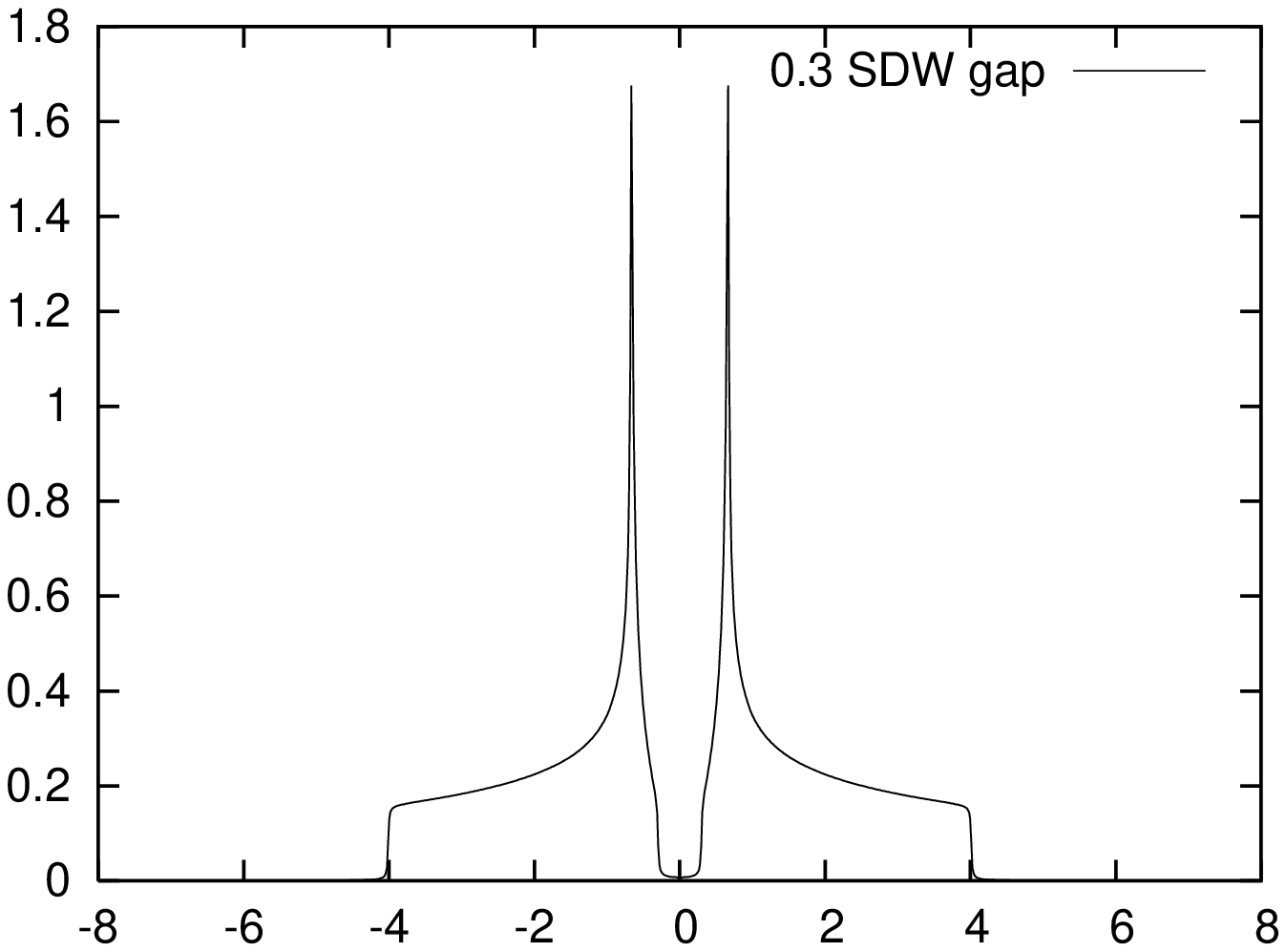}}}\\
      \resizebox{7.8cm}{!}{\rotatebox{270}{\includegraphics{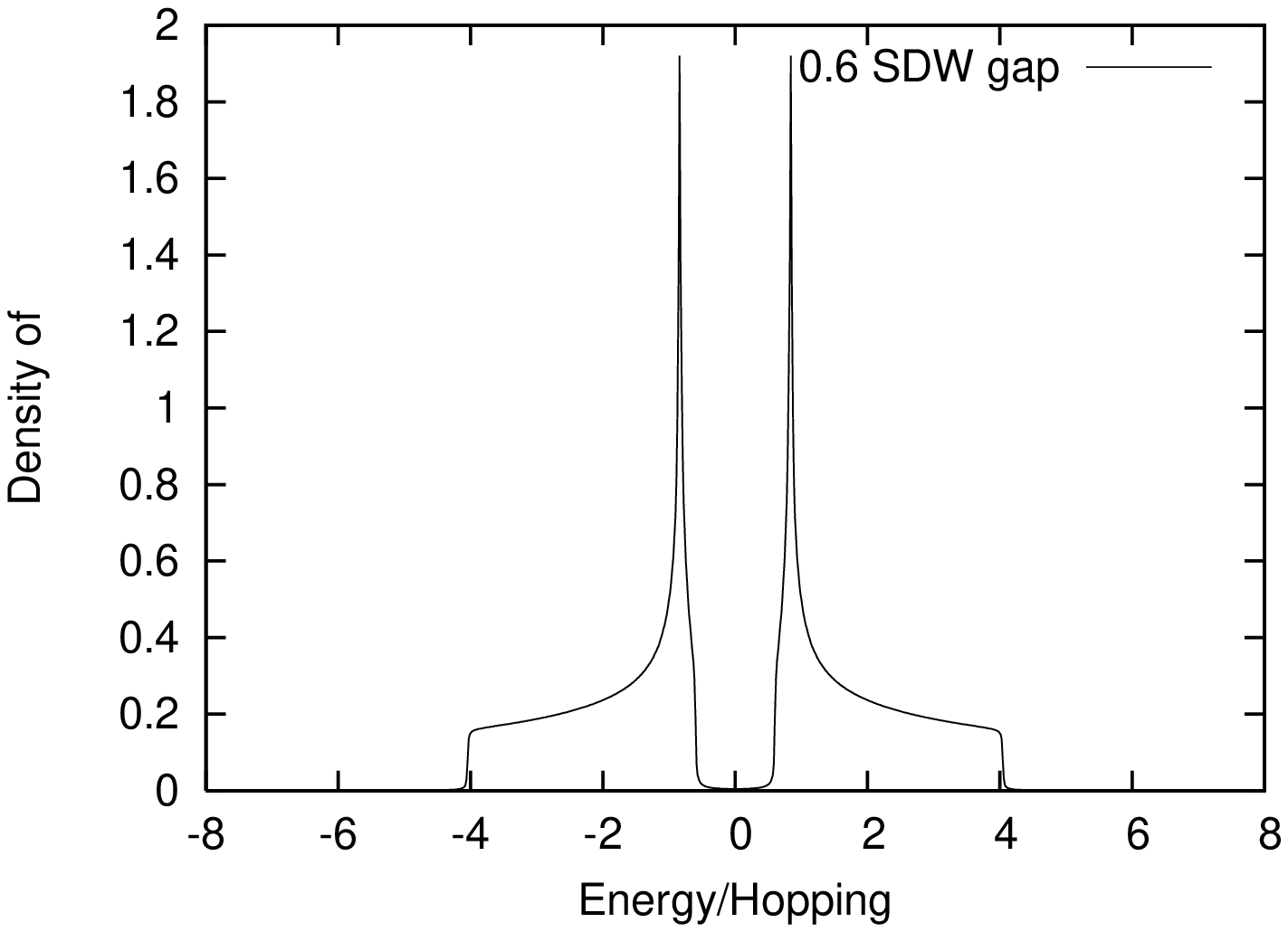}}}&
      \resizebox{7.8cm}{!}{\rotatebox{270}{\includegraphics{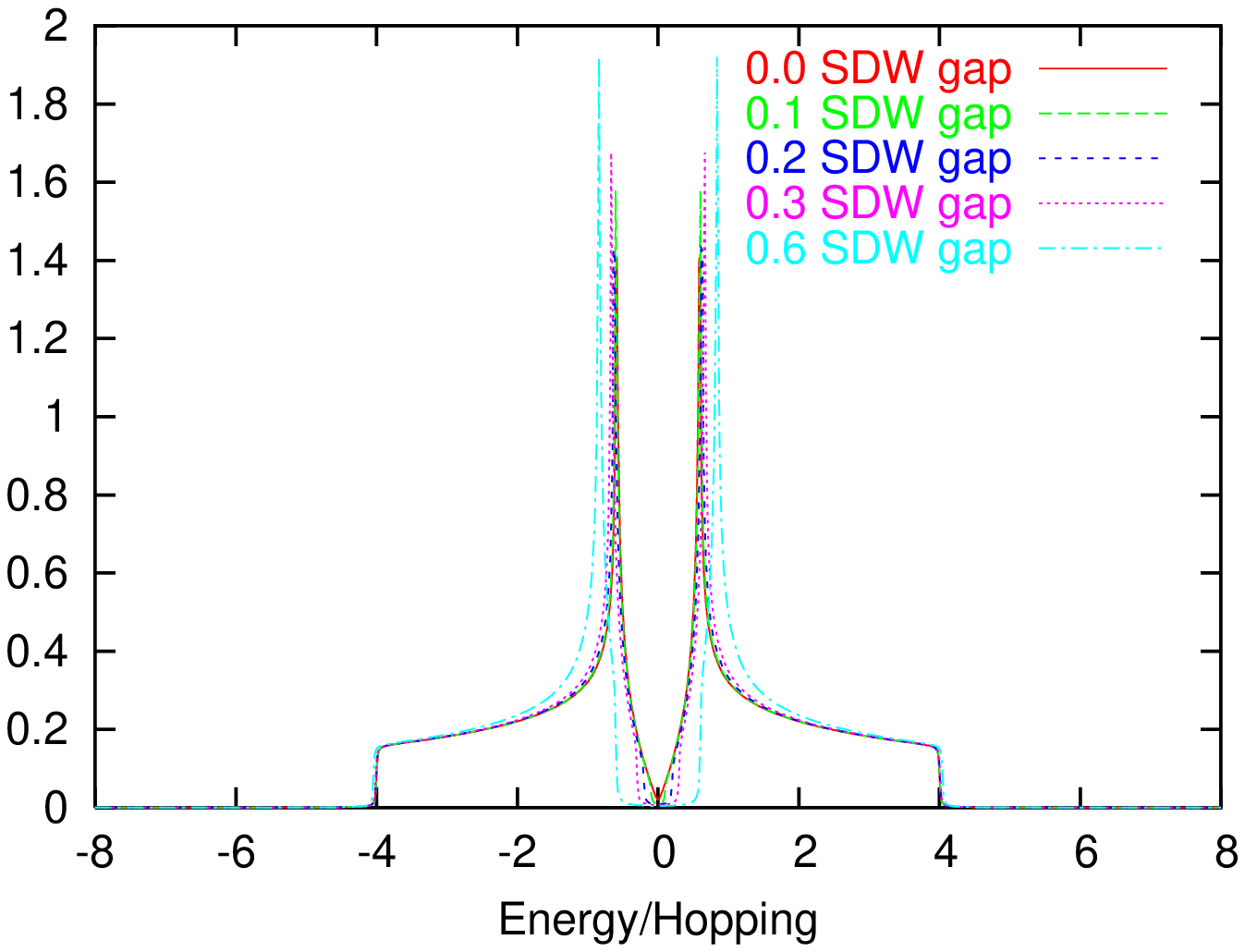}}}
    \end{tabular}
    \caption{Spectral density of states for a d-wave superconductor as
    the SDW gap increases.}
  \end{center}
\end{figure}
\end{widetext}

We will analyze this Hamiltonian by imposing an SDW mean field
condition, which is stabilized by the Hubbard term. This will be
followed by a DSC mean-field condition, which is stabilized by the
reduced BCS d-wave interaction.  While the use of two mean-field
conditions is not common, it has important precedents. It was used by
P. W. Anderson\cite{higgs} in his study of the role of plasmons in
restoring gauge invariance to the BCS ground state. In this work he
invented the Anderson-Higgs mechanism\cite{higgs2}. He solved for the
properties of the electron system imposing a mean-field condition on
the electron density, as in the study of electron correlations by
Sawada, {\it et al}\cite{sawada} and a BCS electron pairing mean-field
condition\cite{bcs}.

The Hubbard interaction stabilizes the mean-field order

\beq \label{selfaf}
\sigma S N \equiv \sum_{\vec{k}} \langle
c_{\vec{k}+\vec{Q},\sigma}^\dagger c_{\vec{k},\sigma} \rangle
\eneq

\noindent where $\vec{Q}=(\pi,\pi)$ is the commensurate ordering wave
vector and $S$ is the average magnetic moment per site. Other ordering
wave vectors are possible for spin and/or charge, i.e. stripe, order
parameters but we do not consider them in our study. When we impose
this condition on the Hamiltonian and neglect fluctuation terms, the
Hamiltonian becomes

\begin{align}
\mathcal{H} = \nonumber \sum_{\vec{k},\sigma} (\epsilon_{\vec{k}} &-
\mu) \; c_{\vec{k},\sigma}^\dagger c_{\vec{k},\sigma} + U N S^2 - U S 
\sum_{\vec{k},\sigma} \sigma c_{\vec{k}+\vec{Q},\sigma}^\dagger
c_{\vec{k},\sigma}\\ 
&+ \sum_{\vec{k}_1,\vec{k}_2} V(\vec{k}_1,\vec{k}_2) \;
c_{\vec{k}_1,\uparrow}^\dagger c_{-\vec{k}_1,\downarrow}^\dagger
c_{-\vec{k}_2,\downarrow} c_{\vec{k}_2,\uparrow}
\end{align}

\noindent We see that by ordering antiferromagnetically we gain
variational energy $- U N S^2$ if self-consistency can be achieved. We
next impose the mean-field d-wave Cooper pairing

\begin{align}
\nonumber \Delta_{\vec{k}_2} &\equiv (cos{k_{2x}}-cos{k_{2y}}) V_0
\sum_{\vec{k}_1} (cos{k_{1x}}-cos{k_{1y}}) \langle
c_{\vec{k}_1,\uparrow}^\dagger c_{-\vec{k}_1,\downarrow}^\dagger
\rangle \\ 
\label{selfsc} &\equiv \Delta_0 (cos{k_{2x}}-cos{k_{2y}})
\end{align}

\noindent Then the Hamiltonian becomes

\begin{align}\label{meanh}
\mathcal{H} = \nonumber \sum_{\vec{k},\sigma} (\epsilon_{\vec{k}} &-
\mu) \; c_{\vec{k},\sigma}^\dagger c_{\vec{k},\sigma} + U N S^2 - U S
\sum_{\vec{k},\sigma} \sigma c_{\vec{k}+\vec{Q},\sigma}^\dagger
c_{\vec{k},\sigma}\\ 
&- \frac{\Delta_0^2}{V_0} + \sum_{\vec{k}} \Delta_{\vec{k}} \;
(c_{\vec{k},\uparrow}^\dagger c_{-\vec{k},\downarrow}^\dagger +
c_{-\vec{k},\downarrow} c_{\vec{k},\uparrow})
\end{align}

\noindent We see that if the phenomenological d-wave interaction is
attractive, i.e. $V_0 < 0$, we gain variational energy $\Delta_0^2/V_0$ by 
Cooper pairing {\it regardless} of whether we have ordered
antiferromagnetically or not. Of course, if we are at half filling,
the material will be insulating {\it irrespective} of the presence of
Cooper pairs, as we would have to excite quasiparticles across the SDW
insulating gap in order for conduction to take place.

\section{Bogolyubov Diagonalization of the Mean-Field Hamiltonian}

We know diagonalize the Hamiltonian by the Bogolyubov
method\cite{bogo1}. We will do this in two steps. First we diagonalize
the SDW part. In order to do this more conveniently, we will split the
momentum sums into sums over the reduced magnetic zone. The
Hamiltonian is then

\begin{align}
\nonumber \mathcal{H} &= \sum_{\vec{k},\sigma}{'} \Big\{
(\epsilon_{\vec{k}}^+ - \mu) (c_{\vec{k},\sigma}^\dagger
c_{\vec{k},\sigma} + c_{\vec{k}+\vec{Q},\sigma}^\dagger
c_{\vec{k}+\vec{Q},\sigma}) \\
\nonumber &+ \epsilon_{\vec{k}}^- (c_{\vec{k},\sigma}^\dagger
c_{\vec{k},\sigma} - c_{\vec{k}+\vec{Q},\sigma}^\dagger
c_{\vec{k}+\vec{Q},\sigma}) - 2 \sigma U S \;
c_{\vec{k}+\vec{Q},\sigma}^\dagger c_{\vec{k},\sigma} \Big\}\\
\nonumber &+ U N S^2 - \frac{\Delta_0^2}{V_0} + \sum_{\vec{k}}{'}
\Delta_{\vec{k}} \; \Big( c_{\vec{k},\uparrow}^\dagger
c_{-\vec{k},\downarrow}^\dagger + c_{-\vec{k},\downarrow}
c_{\vec{k},\uparrow} \\
&- c_{\vec{k}+\vec{Q},\uparrow}^\dagger
c_{-\vec{k}-\vec{Q},\downarrow}^\dagger -
c_{-\vec{k}-\vec{Q},\downarrow} c_{\vec{k}+\vec{Q},\uparrow} \Big)
\end{align}

\noindent where the prime on the summation sign means that the sum is
restricted to the wave vectors in the magnetic
zone. $\epsilon_{\vec{k}}^+ \equiv
(\epsilon_{\vec{k}}+\epsilon_{\vec{k}+\vec{Q}})/2$ and
$\epsilon_{\vec{k}}^- \equiv
(\epsilon_{\vec{k}}-\epsilon_{\vec{k}+\vec{Q}})/2$. The last term in
the superconducting interaction is negative because
$\Delta_{\vec{k}+\vec{Q}}=-\Delta_{\vec{k}}$. In order to diagonalize
the magnetic part we define the Bogolyubov operators

\beq
b_{\vec{k},\sigma} = \alpha_{\vec{k}} c_{\vec{k},\sigma} - \sigma
\beta_{\vec{k}} c_{\vec{k}+\vec{Q},\sigma}
\eneq

\beq
b_{\vec{k}+\vec{Q},\sigma} = \alpha_{\vec{k}}
c_{\vec{k}+\vec{Q},\sigma} + \sigma \beta_{\vec{k}} c_{\vec{k},\sigma}
\eneq

\noindent If we choose 

\beq
\alpha_{\vec{k}}^2 = \frac{1}{2} \left( 1 +
\frac{\epsilon_{\vec{k}}^-}{E_{\vec{k}}} \right) \qquad
\beta_{\vec{k}}^2 = \frac{1}{2} \left( 1 -
\frac{\epsilon_{\vec{k}}^-}{E_{\vec{k}}} \right)
\eneq

\beq
E_{\vec{k}}^2 = (\epsilon_{\vec{k}}^-)^2 + U^2 S^2
\eneq

\noindent the Hamiltonian becomes

\begin{align}
\nonumber \mathcal{H} = \sum_{\vec{k},\sigma}{'} &\Big\{
(\epsilon_{\vec{k}}^+ - \mu) (b_{\vec{k},\sigma}^\dagger
b_{\vec{k},\sigma} + b_{\vec{k}+\vec{Q},\sigma}^\dagger
b_{\vec{k}+\vec{Q},\sigma}) \\
\nonumber &+ E_{\vec{k}} (b_{\vec{k},\sigma}^\dagger
b_{\vec{k},\sigma} - b_{\vec{k}+\vec{Q},\sigma}^\dagger
b_{\vec{k}+\vec{Q},\sigma}) \Big\} + U N S^2 \\
\nonumber &- \frac{\Delta_0^2}{V_0} + \sum_{\vec{k}}{'} \Delta_{\vec{k}} \;
\Big( b_{\vec{k},\uparrow}^\dagger b_{-\vec{k},\downarrow}^\dagger +
b_{-\vec{k},\downarrow} b_{\vec{k},\uparrow} \\
&- b_{\vec{k}+\vec{Q},\uparrow}^\dagger
b_{-\vec{k}-\vec{Q},\downarrow}^\dagger -
b_{-\vec{k}-\vec{Q},\downarrow} b_{\vec{k}+\vec{Q},\uparrow} \Big)
\end{align}

Our last step to diagonalize the full Hamiltonian is the Bogolyubov
diagonalization of the leftover superconducting part by defining the
canonical operators

\beq
B_{\vec{k},\sigma} = u_{\vec{k}}^+ b_{\vec{k},\sigma} + \sigma
v_{\vec{k}}^+ b_{\vec{k},\bar{\sigma}}^\dagger
\eneq

\beq
B_{\vec{k}+\vec{Q},\sigma} = u_{\vec{k}}^- b_{\vec{k}+\vec{Q},\sigma}
- \sigma v_{\vec{k}}^- b_{\vec{k}+\vec{Q},\bar{\sigma}}^\dagger
\eneq

\noindent If we choose

\beq
(u_{\vec{k}}^{\pm})^2 = \frac{1}{2} \left( 1 +
\frac{\epsilon_{\vec{k}}^+ - \mu \pm E_{\vec{k}}}{E_{\vec{k}}^{\pm}}
\right)
\eneq

\beq
(v_{\vec{k}}^{\pm})^2 = \frac{1}{2} \left( 1 -
\frac{\epsilon_{\vec{k}}^+ - \mu \pm E_{\vec{k}}}{E_{\vec{k}}^{\pm}}
\right)
\eneq

\beq
(E_{\vec{k}}^{\pm})^2 = (\epsilon_{\vec{k}}^+ - \mu \pm E_{\vec{k}})^2
+ \Delta_{\vec{k}}^2
\eneq

\noindent the Hamiltonian then becomes

\begin{align}
\nonumber \mathcal{H} = \sum_{\vec{k},\sigma}{'} &\left[ E_{\vec{k}}^+
B_{\vec{k},\sigma}^\dagger B_{\vec{k},\sigma} + E_{\vec{k}}^-
B_{\vec{k}+\vec{Q},\sigma}^\dagger B_{\vec{k}+\vec{Q},\sigma}
\right] \\
&+ U N S^2 - \frac{\Delta_0^2}{V_0} + \text{constants}
\end{align}

\noindent We see that we have two separate superconducting bands with
dispersions $E_{\vec{k}}^+$ and $E_{\vec{k}}^-$. This happens because
the SDW ordering has split the noninteracting band, i.e. the system
with $U=0$. Of course, if the SDW gap were to collapse, the two bands
would merge into one superconducting band. If we look at it from the
opposite perspective, we see that when we turn on the SDW order, there
will be an insulating gap. We have shown in figure 2 how this gap
opens up at the node as calculated in the next section.

The Bogolyubov transformations can of course be inverted to yield the
electron creation and destruction operators in term of the Bogolyubov
eigenoperators ($B_{\vec{k}}$) of the system. From this we can
evaluate the self consistency or ``gap'' equations (\ref{selfaf}) and
(\ref{selfsc}). We obtain

\beq
\frac{2N}{U} = \sum_{\vec{k}} \left( \frac{\epsilon_{\vec{k}}^+ - \mu
+ E_{\vec{k}}}{E_{\vec{k}}^+ E_{\vec{k}}} - \frac{\epsilon_{\vec{k}}^+
- \mu - E_{\vec{k}}}{E_{\vec{k}}^- E_{\vec{k}}} \right)
\eneq

\noindent from the antiferromagnetic self-consistency condition
(\ref{selfaf}) and 

\beq
- \frac{2}{V_0} = \sum_{\vec{k}} (cos{k_x}-cos{k_y})^2 \left(
\frac{1}{E_{\vec{k}}^+} + \frac{1}{E_{\vec{k}}^-} \right)
\eneq

\noindent from the superconducting sel-consistency or gap equation 
(\ref{selfsc}). The
negative sign on the left is consistent with $V_0 < 0$ as is necessary
to stabilize superconductivity. We see that these are two coupled
equations for the superconducting gap parameter $\Delta_0$ and the
spin moment magnitude $S$ in the antiferromagnet. Their solution will
contain information about how the two orders compete and how they rob
spectral weight from each other. In the present work we do not worry
about self-consistency but this is very important and
interesting. Hence it will be part of future work. In this work we
will only concentrate on the spectral properties of the system with
coexisting SDW and DSC correlations.

\section{Green's Function for the SDW-DSC Hamiltonian}

In the present section we will write down the expressions for the
Green's functions for a system with SDW order and d-wave Cooper
pairing in each of the SDW bands. The expression for the retarded
Green's function or the propagator is\cite{fetter}

\begin{align}
\nonumber G(\vec{x},&\vec{x}',t) = \\
\nonumber &-i \sum_{n,\sigma} \Big\{ \; \theta(t)
e^{-iE_nt/\hbar} \langle\psi_0|c_{\vec{x},\sigma}|\psi_n\rangle
\langle\psi_n|c_{\vec{x}',\sigma}^\dagger|\psi_0\rangle \\
\label{green} &+ \theta(-t) e^{iE_nt/\hbar}
\langle\psi_0|c_{\vec{x}',\sigma}^\dagger|\psi_n\rangle
\langle\psi_n|c_{\vec{x},\sigma}|\psi_0\rangle \; \Big\}
\end{align}

\beq 
\theta(t)=-\frac{1}{2\pi i}\int_{-\infty}^\infty
\frac{d\omega}{\omega+i\eta} \; e^{-i\omega t} \qquad \eta=0^+
\eneq

\noindent where

\beq
c_{\vec{x},\sigma}=\frac{1}{\sqrt{N}}\sum_{\vec{k}} c_{\vec{k},\sigma}
e^{-i\vec{k} \cdot \vec{x}}
\eneq

\noindent $n$ labels the eigenstates of the system with energies $E_n$
and the ground state energy $E_0$ has been chosen to be 0.

From the time Fourier transform of the Green's function above we
obtain the local retarded propagator in the energy representation

\begin{align}
\nonumber G(\vec{x},&\vec{x},E) = \frac{1}{\pi N} \sum_{\vec{k}}
\Bigg\{ \frac{(u_{\vec{k}}^+)^2}{E - E_{\vec{k}}^+ + i\eta} +
\frac{(u_{\vec{k}}^-)^2}{E - E_{\vec{k}}^- + i\eta} \\
&-\frac{(v_{\vec{k}}^+)^2}{E + E_{\vec{k}}^+ - i\eta} -
\frac{(v_{\vec{k}}^-)^2}{E + E_{\vec{k}}^- - i\eta} \Bigg\}
\end{align}

\noindent The local spectral density function follows from\cite{fetter}

\beq
A(\vec{x},E)=-\frac{i}{\pi} \; \text{Im} G(\vec{x},\vec{x},E)
\eneq

\noindent All of our density of states are calculated from these
expressions in a $1000 \; \times \; 1000$ momentum lattice with an
energy resolution of 0.01. We choose the hopping energy scale to be 1,
so all energies are measured in hopping units. When we have
superconductivity we choose the gap to be 0.3. The antiferromagnetic
gap is chosen anywhere between 0 and 0.6, usually with jumps of
0.1. We have nearest neighbor hopping only. These values need not be realistic;
they are just chosen to illustrate the effect.

Similarly, if we Fourier transform the Green's function (\ref{green})
in both time and space, we obtain the retarded propagator in the
wavevector energy representation.

\begin{align}
\nonumber G(\vec{k},&E) = \frac{1}{\pi} \Bigg\{
\frac{(u_{\vec{k}}^+)^2}{E - E_{\vec{k}}^+ + i\eta} +
\frac{(u_{\vec{k}}^-)^2}{E - E_{\vec{k}}^- + i\eta} \\
&-\frac{(v_{\vec{k}}^+)^2}{E + E_{\vec{k}}^+ - i\eta} -
\frac{(v_{\vec{k}}^-)^2}{E + E_{\vec{k}}^- - i\eta} \Bigg\}
\end{align}

\noindent From this formula we calculate the absorption strength
vs. energy for the nodal quasiparticles. We do this by simply fixing
$\vec{k}$ to be at the node and plotting the spectral density\cite{fetter}

\beq
A(\vec{k},E)=-\frac{i}{\pi} \; \text{Im} G(\vec{k},E)
\eneq

\noindent vs. energy. Energy units, values and uncertainties are
chosen as described for the local density of states.

\section{Conclusions}

We studied a mean field Hamiltonian with two
mean field order parameters. The Hamiltonian contains a spin-density-wave 
antiferromagnetic mean field stabilized by a Hubbard interaction and a 
d-wave Cooper pairing mean field stabilized by a phenomenological d-wave 
interaction. The two order parameters can coexist and the SDW ground state 
{\it always} gains energy by Cooper pairing when the d-wave interaction is 
attractive and nonzero. The SDW ground state with Cooper pairing {\it fails} 
to superconduct at half-filling due to the antiferromagnetic gap.  Its 
particle-like excitations are Bogolyubov-BCS quasiparticles consisting of
coherent mixtures of electrons and holes. 

Of greater interest and relevance to the superconducting cuprates is
the case when antiferromagnetic order is turned on weakly on top of
the superconductivity. This would correspond to the onset of
antiferromagnetism at a critical doping. In such a case a small gap
proportional to the weak antiferromagnetic gap opens up for nodal
quasiparticles, and the quasiparticle peak would be discernible. While
the gapping of the nodal quasiparticle could be caused by a large enough
disorder, such a disorder would broaden the quasiparticle peak so much
as to make it invisible. A unique signature of antiferromagnetic
gapping of the nodal quasiparticles is that it will turn on always at
the doping when antiferromagnetism starts while disorder gapping will
turn on at different sample dependent dopings.

We wrote down the exact expressions for the Green's function for the system 
with coexisting SDW and DSC order parameters. These are evaluated 
numerically in a $1000 \; \times \; 1000$ momentum lattice with .01 energy 
resolution in units of the lattice hopping. From the imaginary parts of the 
Green's functions we obtained the absorption by nodal quasiparticles and the 
local density of states. 

In our work we did not worry about having self-consistency. This
neglect does not affect our results when the two order parameters are
nonzero, but it will affect whether the order parameters are nonzero
or not, and what the gap values are. Self-consistency will be
important in studying how the two order parameters compete and if and
how they steal spectral weight from each other. Self-consistency might
also affect how the SDW-DSC ground state behaves when doped from
half-filling. Intuitively one expects chemical potential shifts, but
it is not certain that this would be the case. All these issues should
be studied carefully and we postpone them for future work.

{\bf Acknowledgments} Zaira Nazario is a Ford Foundation predoctoral fellow. 
She was supported by the Ford Foundation and by the School of Humanities and 
Science at Stanford University. David I. Santiago  was supported by NASA Grant 
NAS 8-39225 to Gravity Probe B.

\end{document}